\documentclass[%
 reprint,
showpacs,
 amsmath,amssymb,
 aps,prl,
]{revtex4-1}

\usepackage{comment}
\usepackage{graphicx}
\usepackage{dcolumn}
\usepackage{bm,amsmath}
\usepackage{color}

\begin{document}

\preprint{APS/123-QED}

\title{Geometric evolution as a source of discontinuous behavior
       in soft condensed matter}

\author{James E. McClure} 
\affiliation{Virginia Polytechnic Institute \& State University, Blacksburg} 
\author{Steffen Berg}
\affiliation{Shell Global Solutions International B.V.
Grasweg 31,
1031HW Amsterdam,
The Netherlands}
\author{Ryan T. Armstrong}
\affiliation{University of New South Wales, Sydney}

\date{\today}

\begin{abstract}
Geometric evolution represents a fundamental aspect of many physical phenomena. 
In this paper we consider the geometric evolution of structures that undergo
topological changes. Topological changes occur when the shape of an object evolves such 
that it either breaks apart or converges back into itself to form a loop. 
Changes to the topology of an object are fundamentally discrete events.
We consider how discontinuities arise during geometric evolution processes 
by characterizing the possible topological events and
analyzing the associated source terms based on evolution equations for 
geometric invariants. We show that the discrete nature of a topological change 
leads to discontinuous source terms that propagate to physical variables.  

\end{abstract}

\maketitle


Geometric structure frequently influences the behavior of soft matter systems.
Micro-emulsions, liquid crystals, biological tissues, 
and flow through porous media are each governed by closely coupled geometric,
thermodynamic and mechanical processes. The properties and behavior of
soft condensed matter are very much dependent on structure
due to the presence of interfaces, membranes, and other 
energy barriers. 
In this work we focus specifically on interactions that 
result from topological changes, which are alterations to how a structure is connected. 
While the importance of structural effects is frequently obvious, the mathematical
consequences of topological change have been overlooked in the development of physical theory.
Core theoretical results used to formulate arguments about the behavior of
physical systems routinely rely on explicit assumptions of continuity and differentiability \cite{Noether_1918,Noether_1971,Parry_1981,Onsager_1931}. However,
topological changes are not smooth. Understanding how to model the associated structural dynamics 
is critical to develop effective models, both from the geometric perspective and with respect 
to the impact on thermodynamic and mechanical variables.

\begin{figure}[ht]
\centering
\includegraphics[width=1.0\linewidth]{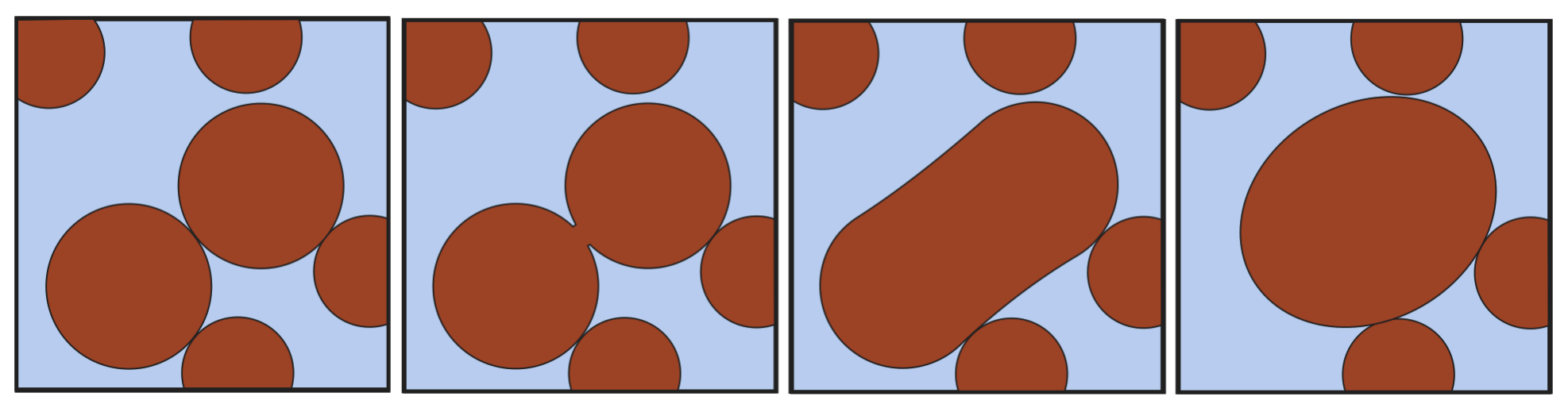}
\caption{Collision-coalescence mechanism within a micro-emulsion. 
After two droplets collide, a bridge forms at the coalescence point.
The larger droplet formed by the event evolves to attain
a new equilibrium configuration based on the minimum potential energy.
}
\label{fig:emulsion-coalescence}
\end{figure}

Discontinuous effects due to topological changes have been noted previously
for fluid systems.
Droplet coalescence involves what is essentially the simplest possible change in connectivity: the merging of two fluid regions. Droplet coalescence is associated with the formation of a singularity, which has been considered in detail from 
an experimental perspective \cite{Pauslon_PRL_2008,Pauslon_PRL_2011,Pauslon_PNAS_2012,Pauslon_NatureComm_2012}. 
Coalescence can dominate the physical behavior from the largest length scales to the smallest, with applications ranging from meteorology to nano-technology, linking with fundamental wetting mechanisms and spreading
\cite{Pak_etal_JPC_2018,Li_etal_nature_2018,Vahabi_etal_science_2018,Berg_2009,Perumananth_etal_PRL_2019}.
The basic collision-coalescence mechanism is illustrated in Fig. \ref{fig:emulsion-coalescence}.
Within emulsions such droplet interactions occur frequently.
In emulsions droplet interactions with films can inhibit coalescence,
implying that the structural properties of the films separating droplets 
are also critical \cite{Zhou_etal_softmatter_2016,Bremond_PRL_2008,Abedi_2019}. 
Such mechanisms are also operative in the behavior of lipid membranes in both biological and engineered systems \cite{Boreyko_etal_pnas_2014}. The effect of fluid structure is 
also evident for flows in porous media, where rapid topological changes that
occur on a fast timescale lead to persistent geometric effects
that can dominate macroscopic behavior \cite{Lenormand_1988,Rucker_2015}.

This paper considers geometric evolution in the general sense, noting that all
connectivity transitions occur as discrete events. At the simplest level these may
be reduced to local coalescence and snap-off phenomena, albeit with 
distinctly more variability. We show that singularities resulting from geometric transitions occur even in well-connected structures, leading to an accumulation of discontinuous 
transitions that influence overall macroscopic behavior. 
An example is considered based on the flow of immiscible fluids
in porous media, demonstrating that snap-off and coalescence events cause 
jump conditions in the time derivative of the fluid pressures. 
Specific contributions of this work are as follows:
(1) characterize geometric evolution processes, showing that in general both continuous and discontinuous transformations must be considered; 
(2) classify the possible topological changes in a three-dimensional system; 
(3) describe how topological change contributes discrete source terms in the fundamental equations that describe geometric evolution;
and (4) demonstrate that geometric discontinuities propagate to physical variables. 

 \section*{Theory}
 
Geometric structure emerges as a consequence of forces acting between
the molecules within a system. Particular molecular configurations lead to local minimum for the potential energy, and the system will spontaneously self-segregate toward such energetically favorable configurations. From the microscopic perspective, the system dynamics are entirely determined from the phase coordinates related to the molecular degrees of freedom. Let us consider a closed system with two distinguishable types of particle, $i\in \{a,b\}$. The classical state of the system is represented based on the position $\bm{q}^{(i)}_k$ and velocity $\bm{\dot{q}}^{(i)}_k$
for each particle, $\bm{r} = \big\{\bm{q}^{(i)}_k,\bm{\dot{q}}^{(i)}_k\big\}$,  
with $k\in\{1,\ldots,N^{(i)}\}$. Geometric structure arises when the molecular system
is treated in an averaged way based on classical thermodynamics and continuum mechanics.

\begin{figure}[ht]
\centering
\includegraphics[width=1.0\linewidth]{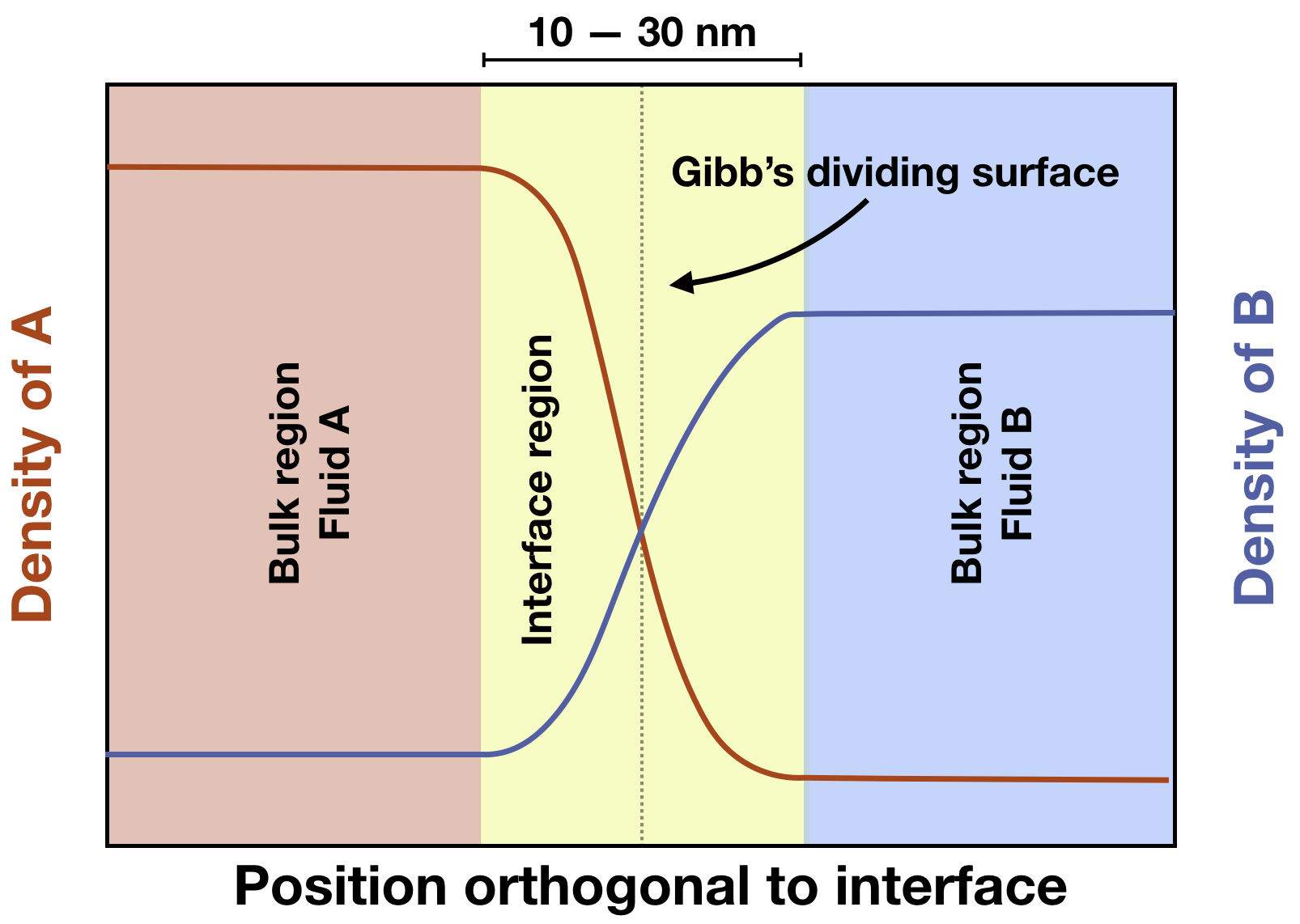}
\caption{The Gibbs dividing surface separates the interface region to obtain
the sharp representation used in classical thermodynamics.
}
\label{fig:gibbs}
\end{figure}

Within the interface region, the composition of the molecular system differs from the bulk phases, as shown in Fig. \ref{fig:gibbs}. The Gibbs dividing surface
is defined to construct particular regions within the system,
allowing separate treatment for three-dimensional phase regions and two-dimensional interface regions \cite{Adamson_Gast_97}.
Since the dividing surface is determined based on the component
densities, the phase regions depend only the position of the particles. 
The two sets formed based on the choice of the Gibbs dividing surface may be 
formally represented as
\begin{equation}
\Omega_i(t) = \Omega_i \big(\bm{q}^{(i)}_k \big) \mbox{for $i \in \{a,b\} $}\; .
\label{eq:subregion-set}
\end{equation}
It follows that invariant geometric measures of $\Omega_i$ depend
only on the particle coordinates. Since the Gibbs dividing surface separates the 
system into a countable number of entities, an inherently discrete element is
introduced into the classical description of the system. 

Sub-division of the system into phase and interface regions is relied upon 
in the classical thermodynamic description of heterogeneous systems \cite{Kjellstrup_Bedeaux_08}. For the
case at hand, the internal energy of the system can be written as
\begin{equation}
    U = \sum_{i \in \{a,b\}} \Big(T_{i} S_i - p_i V_i + \mu_{ik} N_{ik} \Big)
      + \gamma_{ab} A_{ab} \;,
    \label{eq:euler-equation}
\end{equation}
where the relevant quantities for each phase $i\in \{a,b\}$ are
the temperature $T_i$, the entropy $S_i$, the pressure $p_i$,
the volume $V_i$, the chemical potential $\mu_{ik}$ and the number of particles
of type $k$, $N_{ik}$. The energy due to the interfacial arrangement is
given based on the interfacial tension $\gamma_{ab}$ and the associated
surface area separating the two regions. Geometry is therefore introduced
into description of the internal energy of the system. The average pressure within the 
associated phase regions can be determined as
\begin{equation}
    p_i = \frac{ \int_{\Omega_i} p dV}{\int_{\Omega_i} dV} \;.
    \label{eq:pi}
\end{equation}
where the pressure $p$ at any point in the system is defined
based on arguments of statistical mechanics. The spatial average 
given in Eq. \ref{eq:pi} ensures that
the internal energy of the system is properly captured. Analogous arguments
can be made to define spatial averages for other thermodynamic quantities
\cite{Gray_Miller_14}. Due to the structural division of the system based on the Gibb's dividing
surface, each quantity in
Eq. \ref{eq:euler-equation} is either explicitly or implicitly dependent
on the geometric structure of the problem. 

\subsection*{Geometric characterization}

Assumptions regarding the geometric structure of a system provide 
the basis to establish key concepts related to differentiability, symmetry, 
and conservation. Geometric results determined for three-dimensional objects
represent the possible shapes that structures can attain, and derived results also govern how structures can evolve. Hadwiger's characterization theorem establishes that only 
four invariant measures are needed to describe the structure of a three-dimensional object \cite{Hadwiger1957}. Our work proceeds based on these invariant measures. 
A kinematic evolution equation for the volume of a set $\Omega_i$ is 
provided by the Minkowski-Steiner formula. If we consider rolling a sufficiently small ball with diameter $\delta$ around $\Omega_i$, the change in volume is predicted as a linear function of the scalar invariants of the boundary $\Gamma_i$
\begin{equation}
    V(\Omega_i \oplus \delta \zeta) - V(\Omega_i) = \alpha_1 A_i \delta  + \alpha_2 H_i \delta^2 + 
    \alpha_3 \chi_i \delta^3 \;,
    \label{eq:minkowski-steiner}
\end{equation}
where $\zeta$ is the unit ball and the coefficients $\alpha_k$ for $k \in 1,2,3$ are determined based on the structure of $\Omega_i$ \cite{Federer_1959}. 
The scalar invariants are the surface area $A_i$, the mean width $H_i$, and Euler characteristic $\chi_i$. The mean width is defined based on the integral of mean curvature
\begin{equation}
  H_i = \int_{\Gamma_i}  \frac{\kappa_1 + \kappa_2}{2} dS
      \label{eq:Hi}
\end{equation}
where $\kappa_1$ and $\kappa_2$ are the principal curvatures
along the surface. Euler characteristic is directly proportional
to the total curvature, which includes
contributions from the Gaussian curvature of the object
boundary $\Gamma_i$ as well as the geodesic curvature associated
with any non-smooth portions of the boundary $\partial \Gamma_i$,
\begin{equation}
  4\pi \chi_i = \int_{\Gamma_i} \kappa_1 \kappa_2 dS
   + \int_{\partial \Gamma_i} \kappa_g dC
\end{equation}
where $\kappa_g$ is the geodesic curvature. The Euler characteristic 
provides the link to the topology of an object, which is the basis
for using Euler characteristic to measure connectivity \cite{Serra_1983,Nagel_2000}. 
Euler characteristic is related to the alternating sum of the Betti numbers,
\begin{equation}
    \chi_i = B_0 - B_1 + B_2.
    \label{eq:euler}
\end{equation}
where 
$B_0$ is the number of connected components, 
$B_1$ is the number of loops
and $B_2$ is the number of cavities enclosed within the object. 

The Minkowski-Steiner formula is only
valid for sufficiently small values of $\delta$. The coefficients $\alpha_1, \alpha_2$ and $\alpha_3$
are particular functions of $\Omega_i$ and different coefficients may
be obtained for different structures. However, recent work suggests that structures with identical geometric measures will be associated with identical coefficients when 
considered across the state space of many possible geometric structures
 \cite{McClure_Armstrong_etal_2018}.
For additional mathematical background pertaining to the scalar invariants,
the reader is referred to the works of Federer \cite{Federer_1959} and Klain \cite{Klain_95}. Detailed reviews pertaining to the application to characterise the behavior of structures
are also available \cite{Ohser_11,Armstrong_TiPM_2018,Mecke_98,Hilfer_02,Mecke_13}. 
Here we seek to understand how particular structures and their associated invariant measures change with time.

\subsection*{Geometric evolution: a simple example}

Consider first the change in volume that results when growing a sphere with 
radius $r$ to the larger sphere $r+\delta r$
\begin{eqnarray}
V(\Omega_i \oplus \delta r \zeta) -V(\Omega_i) 
&=& \frac {4 \pi }{3} (r+\delta r )^3 - \frac {4 \pi }{3} r ^3 
\nonumber \\
    &=& A_i \delta r  + H_i (\delta r)^2 + \frac {4\pi}{3}\chi_i (\delta r)^3 
\end{eqnarray}
where expressions for the geometric invariants associated with a sphere have
been inserted: $A_i=4\pi r^2$, $H_i=4\pi r$ and $\chi_i=1$.
Comparing with Eq.\ref{eq:minkowski-steiner} we quickly see that the coefficients are 
$\alpha_1 = 1$, $\alpha_2 = 1$ and $\alpha_3 = 4\pi/3$. Next we consider 
a torus with major radius $R$ and minor radius $r$
 \begin{eqnarray}
V(\Omega_i \oplus \delta r \zeta) -V(\Omega_i) &=&  2 \pi^2 (r+\delta r )^2 R - 2 \pi^2 r^2 R \nonumber \\
    &=&  A_i \delta r + H_i (\delta r)^2 
\end{eqnarray}
using the fact that for the torus $A_i=4\pi^2 rR$, $H_i=2 \pi^2 R $ and $\chi_i=0$. Again referring to 
Eq.\ref{eq:minkowski-steiner}, we see that $\alpha_1 = 1$ and $\alpha_2 = 1$. Even though $\chi_i=0$ for a torus, identical coefficients are obtained for the two structures.

\begin{figure}[ht]
\centering
\includegraphics[width=0.8\linewidth]{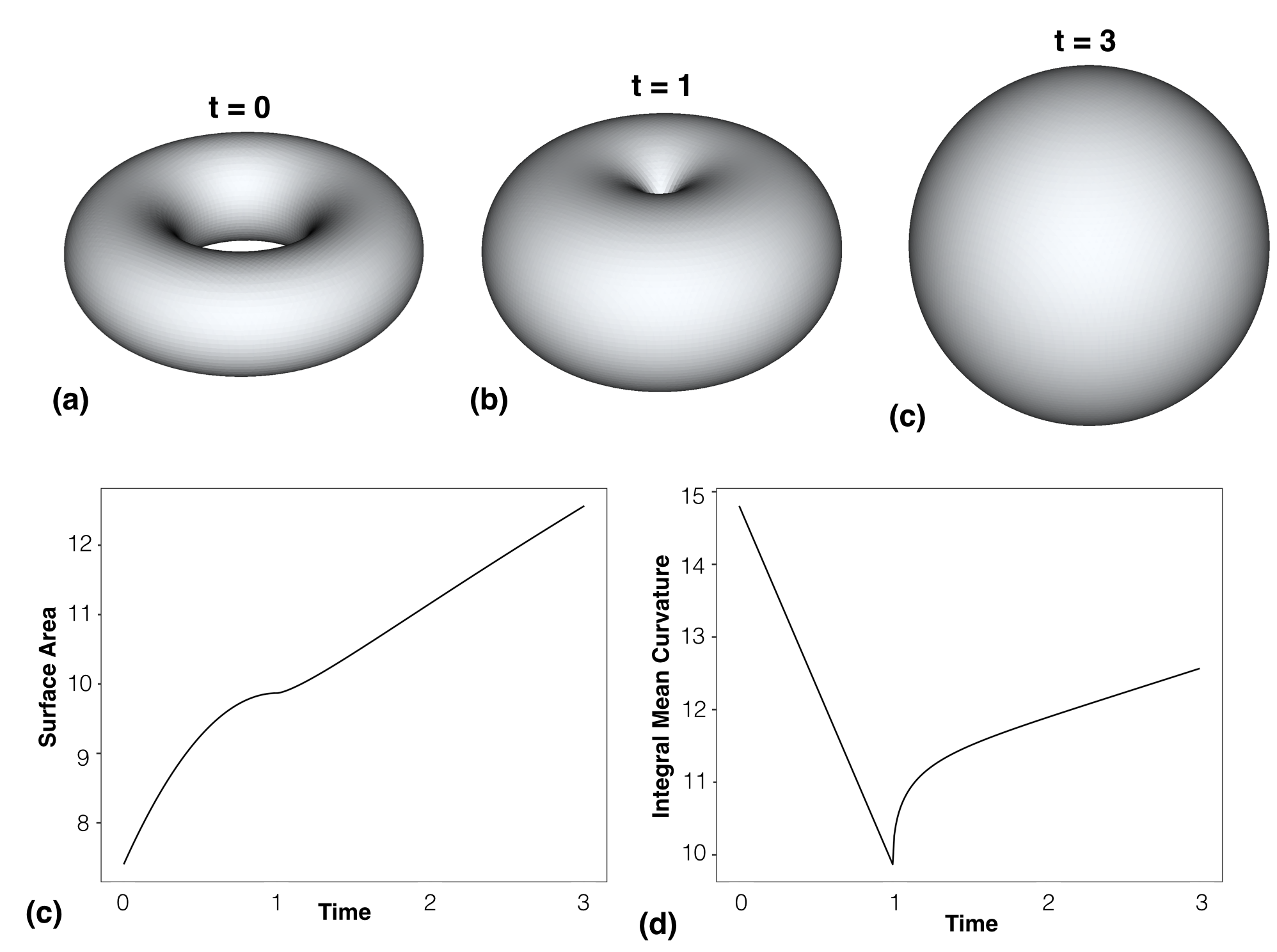}
\caption{The evolution of a torus into a sphere.
The center hole of the torus closes at $t=1$ causing a jump
condition in the Euler characteristic.
Plots showing the (c) surface area 
and (d) mean width as a function of time based on 
the evolution of a torus into a sphere. The hole in
the center of the torus closes at time $t=1$, resulting in time
dependence that is non-smooth. Changes in connectivity
are linked to a breakdown in local smoothness, presenting a fundamental
challenge to modeling the geometric evolution and associated physics.}
\label{fig:torus-evolution}
\end{figure}

An important challenge associated with modeling geometric evolution
is that changes in the connectivity of an object occur as discrete events; 
$\chi_i(t)$ is not a continuous function. This means
that while the volume, surface area and mean width will necessarily be continuous functions, the Euler characteristic will evolve based on a series of discrete jumps.
To illustrate this behavior we consider the geometric evolution of a torus into
a sphere as shown in Fig. \ref{fig:torus-evolution}. 
This relatively simple problem is significant because it defines
a change in connectivity. The initial torus has
$\chi=0$; for the final sphere  $\chi=1$. Based on Eq. \ref{eq:euler} 
we can see that for the torus $B_0 = 1$, $B_1 = 1$ and $B_2=0$.
The single handle disappears at time $t=1$ and $B_0 = 1$, $B_1 = 0$ and $B_2=0$. 
The change in connectivity occurs at the instant the hole in the center
of the torus closes, representing a discontinuity in the geometric evolution
process. We now examine the ramifications for this change. Specifically, we will show 
that the discontinuity in the time evolution for the Euler characteristic leads to non-smooth behavior for the other geometric invariants.

The torus is defined based on a surface of revolution for a circle
with minor radius $r$ that is displaced from the origin based on the major radius $R$. 
We choose initial minor radius $r_0= 1/4$ with $r$ an increasing function of time,
\begin{equation}
    r(t) = r_0 + \frac{t}{4} \;.
\end{equation}
We enforce $r+R =1$ such that the object has constant unit width.
For time $t<1$ the circle has radius $r < R$ and the surface of revolution 
is a torus. At time $t=1$ the circle has radius $r=R=1/2$ and the surface 
becomes a spindle torus for $1<r<3$. At $t=3$ the object is 
the unit sphere. The associated three-dimensional structures
are shown in Fig. \ref{fig:torus-evolution}.

To compute the geometric invariants, we rely on expressions derived for
surfaces of revolution. The circle is parameterized by 
\begin{eqnarray}
x(s) &=& R + r \cos (s) \\
y(s) &=& r \sin (s).
\end{eqnarray}
Since the surface of revolution is self-intersecting for $r>R$, the limits of integration are defined based on the angle
\begin{equation}
\alpha = \left \{ 
\begin{array}{cl}
    \arccos (R / r) & \mbox{if $r>R$ }  \\
    0 & \mbox{otherwise.} 
\end{array} \right.
\end{equation}
The principle curvatures are
\begin{eqnarray}
\kappa_1 &=& \frac{1}{(\dot{x}^2 + \dot{y}^2)^{3/2} } 
\big( -\ddot{x}  \dot{y} + \dot{x} \ddot{y} \big)\;,  \\
\kappa_2 &=& \frac{1}{(\dot{x}^2 + \dot{y}^2)^{1/2} } \frac{\dot{y}}{x} \;.  
\end{eqnarray}
The surface area is 
\begin{eqnarray}
A(t) &=& 2 \pi \int_{-\pi + \alpha}^{\pi - \alpha} 
           x(s) \sqrt{\dot{x}^2 + \dot{y}^2} ds \nonumber \\
  &=& 2 \pi [Rr (2 \pi - 2\alpha) +2r^2 \sin (\alpha)]
  \label{eq:A(t)}
\end{eqnarray}
For time $t<1$, $\alpha=0$ and the expression reduces to $A=4\pi Rr$. For time $t=3$,
the expression for a sphere is obtained, $A=4\pi r^2$, since $R=0$. 
Based on a similar calculation we can calculate the mean
width
\begin{eqnarray}
H(t) &=& \pi \int_{-\pi + \alpha}^{\pi - \alpha} 
           (\kappa_1 + \kappa_2) x \sqrt{\dot{x}^2 + \dot{y}^2} ds\nonumber \\
    &=& 2 \pi R (\pi -  \alpha ) + 4 \pi r \sin(\alpha)\;.
\label{eq:H(t)}
\end{eqnarray}
Eqs. \ref{eq:A(t)} and \ref{eq:H(t)} are plotted in Fig. \ref{fig:torus-evolution}, clearly showing that while
$A(t)$ and $H(t)$ are both continuous functions, the time derivatives 
are discontinuous at the instant that
the hole in the middle of the torus closes ($t=1$), which is also associated with
a jump condition in the Euler characteristic from $\chi=0$ to $\chi=1$. 
Predicting when such topological changes occur requires information regarding
the overall geometric structure of an object. 

Due to the topological change, a one-to-one mapping cannot be defined
for the geometric evolution at $t=1$. Prior to the topological change 
we can identify the ring 
of points at the interior boundary of the torus $x,y: x^2+y^2-(R-r)^2 = 0; z=0$.  
At time $t=1$ the ring closes such that the entire ring of points are mapped to the 
origin, which means that the associated mapping is not one-to-one. The geometric evolution 
can be considered based on a continuous deformation of the structure for $t<1$ 
and then again for $t>1$, but
not for $t=1$. This underscores the fact that geometric evolution should 
be generally considered as a sequence of processes divided into two distinct categories:
\begin{enumerate}
\item Continuous local geometric flow at constant topology as given by 
the time dependent mapping $G_k^t: \Omega \rightarrow \Omega$
where $G_k^t$ is one-to-one for time $t_{k-1} < t < t_k$.
\item Discrete geometric transformations defined by 
$G_k^*: \Omega \rightarrow \Omega$
occurring at time $t_k$ and corresponding to a topological change that cannot be represented 
based on a bijection.
\end{enumerate}
By chaining together a sequence of such mappings $k\in\{1,2,\ldots,K\}$, 
it is possible to model geometric evolution processes of arbitrary complexity. 

\subsection*{Classification of Topological Changes}

\begin{figure}[ht]
\centering
\includegraphics[width=1.0\linewidth]{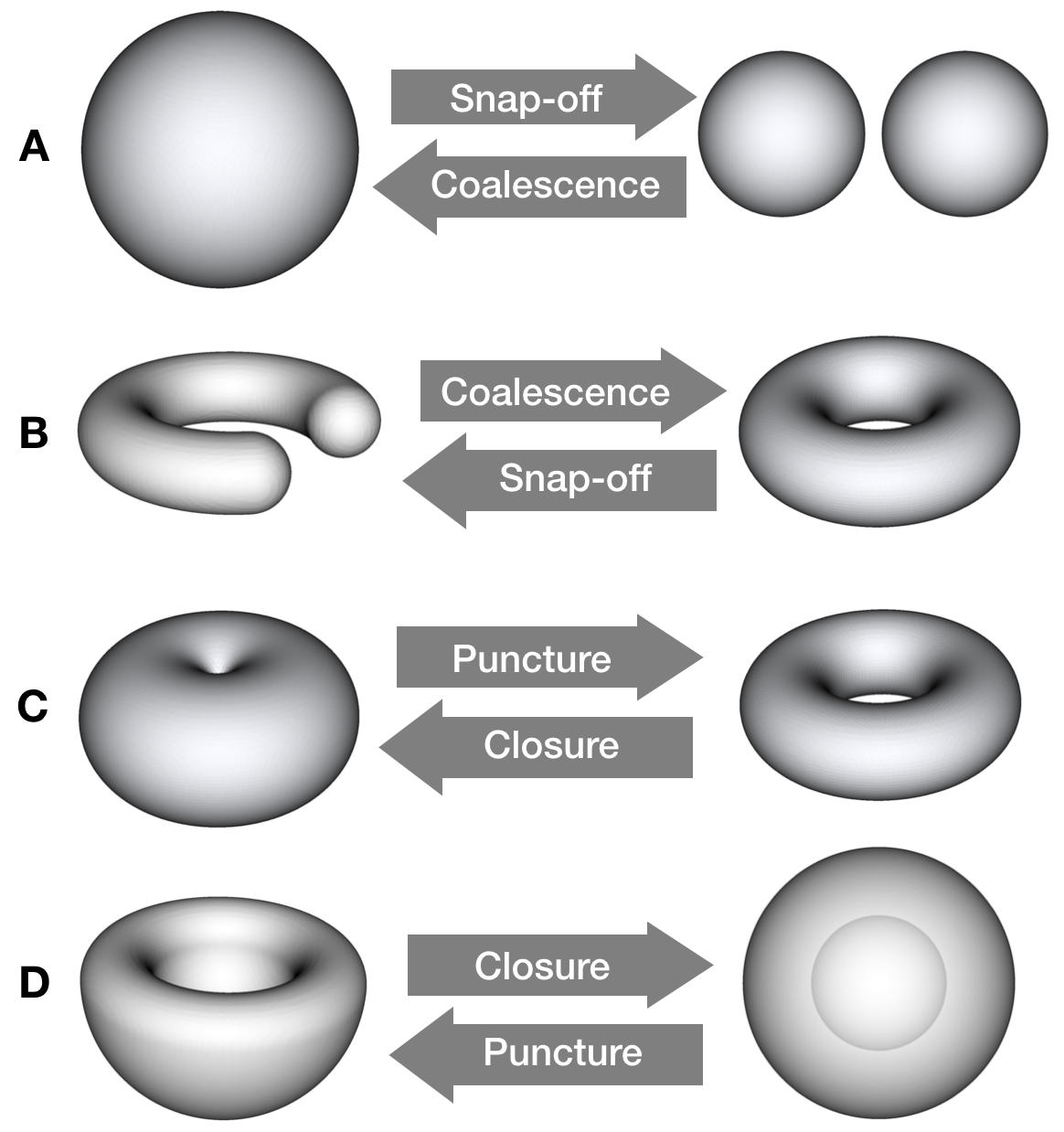}
\caption{Discontinuous geometric maps in three-dimensions fall into
eight categories based on homeomorphism. The associated topological
changes increase or decrease Euler characteristic by exactly one.}
\label{fig:discontinuous-maps}
\end{figure}

Only a limited number of topologically distinct
structures are possible within a particular dimensional space \cite{Thurston_97}.
In a three-dimensional system, all possible topological states
can be reduced to the analysis of three fundamental shapes: the sphere, the torus, and the spherical shell. 
These shapes are linked to the Betti numbers: the sphere is a single
connected component and links to $B_0$; the torus has a single loop
and links to $B_1$;  the spherical shell contains a single cavity
and links to $B_2$. To form more complex structures, these shapes 
can be placed in a system, first gluing objects together
as needed, then stretching and deforming the resulting object
until a desired structure is obtained. Based on this it is possible to 
characterize  the possible topological changes that may occur during 
geometric evolution, i.e. the discontinuous maps $G^*_k$. 
These eight possibilities are depicted in Fig. \ref{fig:discontinuous-maps}.

Objects on left side of Fig. \ref{fig:discontinuous-maps} can be produced
by continuously stretching and deforming a a sphere (i.e. homeomorphisms). 
The depicted topological changes increase or decrease 
the Euler characteristic by exactly one. Moving left to right for case 
A, the snap-off mechanism generates two regions homeomorphic to a sphere
from a single such region. The consequence is to increase 
Euler characteristic by one due to the corresponding 
increase in $B_0$. The coalescence mechanism corresponds to the
opposite situation, where two regions merge together to destroy
one connected component. The pair of transformations labeled as B rely 
on the same underlying mechanism, in this case either forming or destroying a loop. 
Moving left to right, the Euler characteristic decreases by 
one due to a corresponding increase to $B_1$. Snap-off has the opposite effect.

For four transformations given by A and B the singularities are isolated to particular points, 
which represent the sub-regions of the set for which the mapping is not bijective. For the four transformations given by C and D, the singularity involves a ring structure. For the pair of transformations labeled as C, the puncture mechanism forms a ring of 
points at the location of the singularity. This transformation increases $B_1$ 
and decreases $\chi$. The closure mechanism involves the collapse of
a ring of points to the singularity, destroying the loop. This is the evolution considered in Fig. \ref{fig:torus-evolution}. The pair of transformations shown in Fig. \ref{fig:discontinuous-maps}D shows that the closure mechanism can also form a
cavity starting from a bowl-shaped region. This increases both $B_2$ and 
$\chi$ by one. The puncture mechanism can re-open the hole to destroy the cavity. 
For each of the eight cases, the associated mappings are not bijective. Noting
that a single object can undergo multiple such transformations, 
we can always identify a neighborhood of points that surround the 
singular point, effectively isolating each discontinuity in space and time.

\subsection{Geometric evolution: a hierarchical perspective}

One can further gain insight into the nature of geometric discontinuities 
by considering the time evolution based on a hierarchical view of Eq. \ref{eq:minkowski-steiner}. 
Given an arbitrary closed, three-dimensional object
we consider continuous changes in volume such that the first derivative can be 
defined with respect to time. From Eq. \ref{eq:minkowski-steiner} we can easily see that if 
$\delta$ is sufficiently small
\begin{equation}
    V(\Omega_i \oplus \delta \zeta) - V(\Omega_i) = \alpha_1 A_i (t) \delta + \mathcal{O}(\delta^2)
    \label{eq:MS-hierarchy-Vi} \;.
\end{equation}
That is, the infinitesimal change in volume is entirely given by the movement of the boundary,
and the surface area is the only boundary invariant on which the volume change depends. 
The same insight can be obtained by applying the Reynolds transport theorem to the
region $\Omega_i$ to predict the time rate of change
\begin{equation}
\frac{d V_i}{d t} = \int_{\Gamma_i} (\bm{w}_i \cdot \bm{n}_i) dS \;,
\label{eq:Reynolds}
\end{equation}
where $\bm{n}_i$ is the outward normal to the boundary and $\bm{w}_i$
is the boundary velocity. The time derivative for the surface area and mean width can also be identified from boundary boundary integrals. Since the change in surface area due to the deformation of a local 
surface element is determined by the curvature of the surface element \cite{Gray_Leijnse_1993},
\begin{equation}
\frac{d A_i}{d t} =  \int_{\Gamma_i} \frac{\kappa_1 + \kappa_2}{2}  
(\bm{w}_i \cdot \bm{n}_i) dS \;,
\label{eq:Reynolds-Ai} 
\end{equation}
and similarly for the mean curvature,
\begin{equation}
\frac{d H_i}{d t} = 2 \int_{\Gamma_i} {\kappa_1 \kappa_2} 
(\bm{w}_i \cdot \bm{n}_i) dS \; .
\label{eq:Reynolds-Hi}
\end{equation}
Eqs. \ref{eq:Reynolds} -- \ref{eq:Reynolds-Hi}
describe how the geometric invariants evolve during the continuous
portions of the geometric evolution of an object. The integrals capture both
the growth and the deformation of the object boundary. A topological source
term can be identified in Eq. \ref{eq:Reynolds-Hi} due to the 
dependence on the Gaussian curvature. 

Noting that a differential equation cannot be derived for Euler characteristic, 
it is useful to predict the topological state based on the geometric state function
\begin{equation}
    \chi_i = \chi(V_i,A_i,H_i)\;.
    \label{eq:topological-state}
\end{equation}
Previous work indicates that the geometric state can be predicted uniquely
for complex structures based on a relationship between the four geometric invariants
\cite{McClure_Armstrong_etal_18}. In the context of geometric evolution,
expressing the relationship as a predictor of the topological state is particularly useful;
together with Eqs. \ref{eq:Reynolds}--\ref{eq:Reynolds-Hi}, an equation is
associated with each of the four geometric invariants. While the time evolution is 
given generally from integral equations, approximate forms have also been explored 
as a way to apply the result to particular physical systems \cite{Gray_Dye_etal_15}.

An intuitive basis for how topological source terms arise within
geometric evolution can be established based on further consideration
of the Minkowski-Steiner formula.
Based on Eq. \ref{eq:Reynolds} it is natural to define the average boundary
displacement as
\begin{equation}
\xi \equiv \frac{\int_{\Gamma_i} (\bm{w}_i \cdot \bm{n}_i) dS}{A_i} \;.
\label{eq:displacement}
\end{equation}
For the special case where the boundary velocity in
the normal direction is constant on $\Gamma_i$, $\xi = \bm{w}_i \cdot \bm{n}_i$ and we can make a direct link with Eq. \ref{eq:MS-hierarchy-Vi}. For a volume change over some time $\Delta t$ it is evident that $\delta = \xi \Delta t$ and $\alpha_i=1$, meaning that
Eqs. \ref{eq:Reynolds} -- \ref{eq:Reynolds-Hi} may be expressed as
\begin{eqnarray}
\frac{d V_i}{d t} &=& A_i(t) \xi ,
\label{eq:MS-hierarchy-1} \\
\frac{d A_i}{d t} &=& \alpha_2 H_i(t) \xi ,
\label{eq:MS-hierarchy-2} \\
\frac{d H_i}{d t} &=& \frac{\alpha_3}{\alpha_2} \chi_i(t) \xi .
\label{eq:MS-hierarchy-3}
\end{eqnarray}
That is, to predict the time evolution we need to know the coefficients $\alpha_2$
and $\alpha_3$ as well as the Euler characteristic $\chi_i$. Based on these arguments it is intuitive to express the Minkowski-Steiner formula in a hierarchical manner,
\begin{equation}
    V(\Omega_i \oplus \delta \zeta) - V(\Omega_i) =  \Big[ A_i  + \alpha_2 \big(H_i + 
    \frac{\alpha_3}{\alpha_2 } \chi_i \delta \big)\delta \Big] \delta \;.
    \label{eq:minkowski-steiner-2}
\end{equation}

Eqs. \ref{eq:MS-hierarchy-1}-- \ref{eq:MS-hierarchy-3} offer insight into how the jump condition due to change in Euler characteristic propagates to other variables. From Eq. \ref{eq:MS-hierarchy-3} we can deduce that
$H_i(t)$ is not differentiable whenever the topology of the system changes,
which is implied by the dependence on $\chi_i$.
$A_i(t)$ inherits a discontinuity in the second-order time derivative from
$H_i$. Both results are confirmed based on inspection of
Fig. \ref{fig:torus-evolution}. It can also 
be seen that since the spindle torus does not satisfy the requirement of
positive reach, $\alpha_2$ and $\alpha_3$ are time-dependent. However, 
this apparently does not alter the associated differentiability class for
$H_i$ or $A_i$. Of course, more favorable smoothness properties should be expected
in the regions of the time where no topological changes occur.

Discontinuous aspects of geometric evolution represent a critical challenge to modeling
physical behavior. Systems that undergo topological change will involve
non-smooth aspects with respect to the time evolution for the geometric invariants.
An essential question is the extent to which geometric effects impact continuity and
differentiability for thermodynamic and mechanical variables. Noether's theorem provides 
a formal basis for the continuous description of physical systems based on differential equations \cite{Noether_1918,Noether_1971}. Noether's theorem focuses on invariant properties for systems that undergo continuous group transformations (i.e. Lie groups), relying explicitly 
on invertibility. Typical proofs of the Poincar\'{e}
recurrence relation also rely on the invertability of group transformations \cite{Parry_1981}. 
The continuous transformations $G_k^t$ fall within this category. These correspond to continuous 
deformation of an object at constant topology, which proceed during
a finite interval of time. Topological changes are associated with
a discontinuous map $G_k^*$, which fall into a class of transformations
that are explicitly excluded from Noether's arguments. The implication is that 
even though Noether's arguments apply to the molecular system, the choice to represent
the behavior of a heterogeneous system based on the classical thermodynamic description
will effectively destroy the underlying symmetry within the defined regions. 
Particular consequences become evident when
non-stationary processes are considered. Differentiability of the terms that appear in 
Eq. \ref{eq:euler-equation} is required to develop essential
results of non-equilibrium thermodynamics, which in turn play a major role in
the description of soft matter systems \cite{deGroot_Mazur_84}. Topological
effects must be treated carefully, since the associated discontinuities
undermine the symmetry arguments that form the basis for continuum-scale descriptions.

\section*{Results}

The flow of immiscible fluids in porous media are strongly influenced by geometric and topological effects \cite{Land_68,Lenhard_Parker_87,Herring_Harper_etal_13,McClure_Berrill_etal_16b,Liu_Herring_etal_17,McClure_Armstrong_etal_18,Schlueter_Berg_etal_15,Hilfer_06,Armstrong_McClure_etal_16}. Capillary forces
typically dominate such flows, and these forces are directly impacted by
changes to fluid connectivity. 
Coalescence events always involve the local destruction of interfacial 
curvature, which instantaneously alters accompanying capillary forces. The
Laplace equation relates the pressure difference between
of the adjoining fluids and the capillary pressure given by the product of 
the interfacial tension $\gamma$ and mean curvature:
\begin{equation}
    p_n - p_w = \frac{\gamma}{2} \Big(\kappa_1 + \kappa_2 \Big) \;.
    \label{eq:laplace}
\end{equation}
The Laplace equation holds only at mechanical equilibrium, and a rapid 
change to the interface mean curvature will cause an immediate force imbalance.
The associated discontinuity is directly evident from Eq. \ref{eq:Hi}
and the associated dynamics. The coalescence event illustrated in
Fig. \ref{fig:emulsion-coalescence} provides an intuitive example.
Immediately prior to a coalescence event, the interface
curvature is positive based on the outward pointing normal vector relative to the 
object boundary. Coalescence leads to the formation of a bridge, which causes the sign of the curvature 
to change from positive to negative. The local disruption to the geometric structure 
creates a force imbalance that is analogous to puncturing a balloon with a pin;
coalescence effectively ruptures the interface. The local fluid pressures must 
respond rapidly to correct the force imbalance. However, the timescale for this to
occur is coupled to geometric evolution. Mechanical equilibrium cannot be
re-established until the interfaces attain an appropriate shape. Detailed studies
of the coalescence mechanism show that the flow behavior and geometric evolution
are coupled on a very short timescale \cite{Pauslon_PRL_2011}.

We consider fluid displacement within a porous medium as a means to explore how geometric 
changes influence physical behavior. A periodic sphere packing
was used as the flow domain. Using a collective rearrangement algorithm, 
eight equally-sized spheres with diameter $D=0.52$ mm 
were packed into a fully-periodic domain $1 \times 1 \times 1$ mm in size.
The system was discretized to obtain a regular lattice with $256^3$ voxels. 
Simulations of fluid displacement were performed by injecting fluid into the system
so that the effect of topological changes could be considered in detail. 
Initially the system was saturated with water, and fluid displacement was then simulated based 
on a multi-relaxation time (MRT) implementation of the color lattice Boltzmann model \cite{McClure_Prins_etal_14}.

\begin{figure}[ht]
\centering
\includegraphics[width=1.0\linewidth]{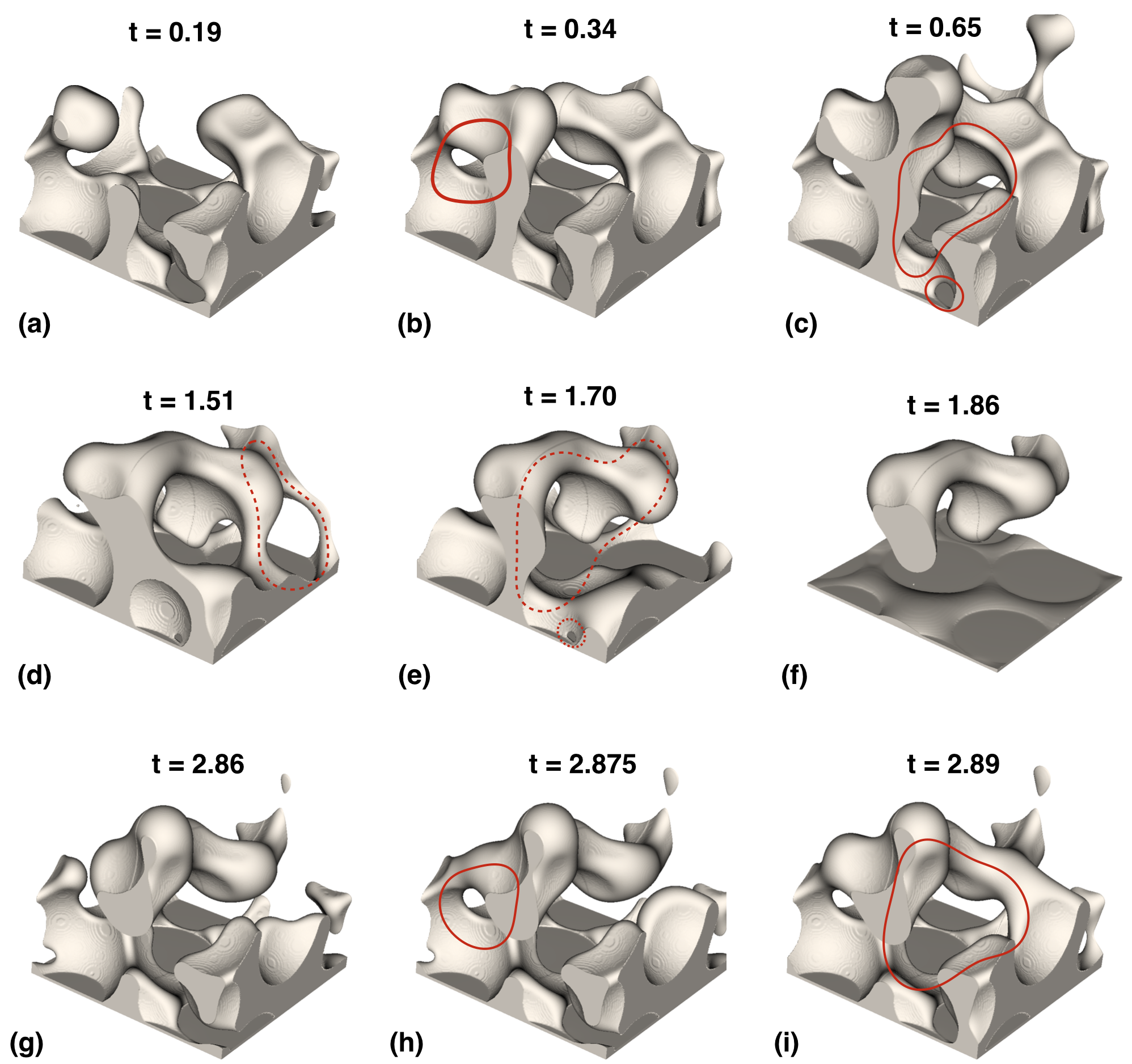}
\caption{Coalescence and snap-off events alter the topology
of fluids within porous media:
(a)-(c) non-wetting fluid loops formed during drainage lead to a decrease in Euler characteristic;
(d)-(f) imibition destroys fluid loops due to snap-off, eventually trapping non-wetting fluid within the porespace; 
(h)-(i) trapped fluids reconnect during secondary drainage. 
}
\label{fig:porescale}
\end{figure}

To simulate primary drainage, non-wetting fluid
was injected into the system at a rate of $Q=50$ voxels per timestep,
corresponding to a capillary number of $4.7\times 10^{-4}$.  A total of
500,000 timesteps were performed to allow fluid to invade the pores within the system, forming the sequence of structures shown in Fig. \ref{fig:porescale} a--c. The first pores are invaded at $t=0.19$,
which is just after the system overcomes the entry pressure. 
The fluid does not yet form any loops, since 
a pore must be invaded from two different directions before a loop will
be created. Haines jumps and loop creation can be considered as
distinct pore-scale events that may coincide in certain situations.
At time $t=0.34$ the first loop is formed. As primary drainage progresses 
additional loops form such that the total curvature tends to decrease as 
non-wetting fluid is injected.

At the end of primary drainage the flow direction was reversed to induce imbibition, with
water injected at a rate of $Q = 20$ voxels per timestep. Sequences from the displacement
are shown in Fig. \ref{fig:porescale} d--f. As water imbibes into the
smaller pores in the system, there is a well-known tendency to snap off non-wetting fluid between connections to larger adjacent pores where imbibition has not yet occurred.
This is known as the Roof snap-off mechanism \cite{Roof_1970}. Snap-off breaks loops
within the system, causing the Euler characteristic to increase. The order that loops
snap off during imibition is not the reverse of the order that the loops form during drainage. This is because the drainage process is dominated by the size of the throats whereas imibition is governed by the size of the pores. Eventually, 
the sequence of snap-off events can cause non-wetting fluid to become disconnected
from the main region, leading to trapping as seen at time $t=1.86$.


The presence of trapped fluid distinguishes secondary drainage from 
primary drainage. The sequence of
fluid configurations shown in Fig. \ref{fig:porescale} depicts 
the sequence of coalescence events that reconnect trapped fluid. 
Due to the presence of trapped fluid, the order that loops are formed
within the porespace does not match the order observed during primary 
drainage. 

\begin{figure}[ht]
\centering
\includegraphics[width=1.0\linewidth]{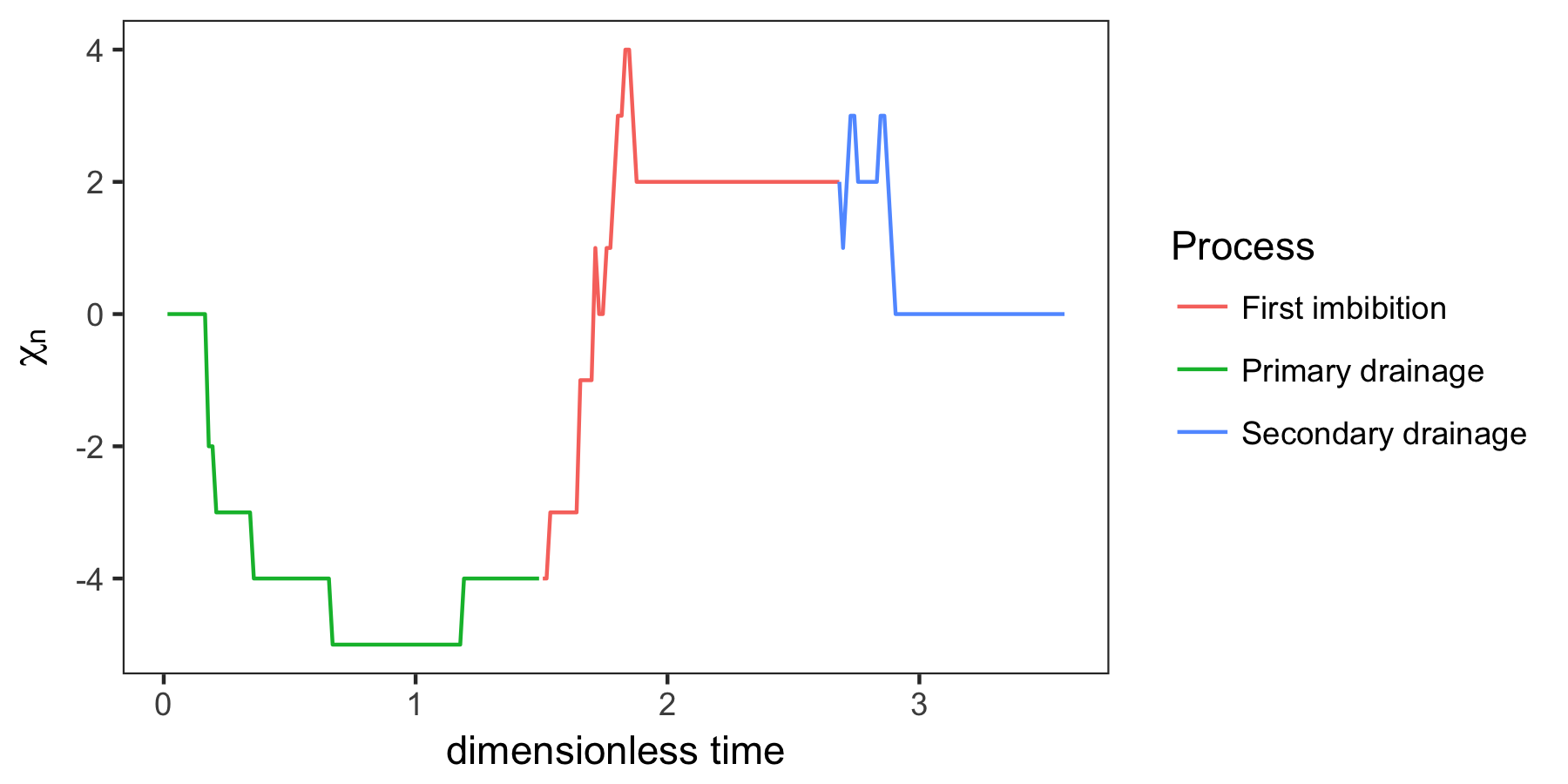}
\caption{Euler characteristic evolves based on a sequence of jump conditions
that correspond to connectivity events within the porespace.  Euler characteristic decreases due to coalescence events and increases due to snap-off. }
\label{fig:Xn}
\end{figure}

Structural changes that occur during displacement are quantitatively 
captured based on the evolution of the Euler characteristic. As shown in
Fig. \ref{fig:Xn}, Euler
characteristic always takes on an integer value due to the 
relationship to number of loops and disconnected regions -- it is not
possible to create a partial loop. The Euler characteristic consequently
evolves based on a sequence of discrete jump conditions that align precisely
with the connectivity events shown in in Fig. \ref{fig:porescale} a-i. As loops form
during drainage, Euler characteristic decreases. As loops are destroyed
during imbibition, Euler characteristic increases. 
We see clearly from Fig. \ref{fig:Xn} that the evolution of the Euler characteristic
can be considered as piece-wise continuous, with the continuous portions being
constant. Predicting the evolution for the Euler characteristic is therefore entirely a question of being able to predict when the jump conditions will occur. 


\begin{figure}[ht]
\centering
\includegraphics[width=1.0\linewidth]{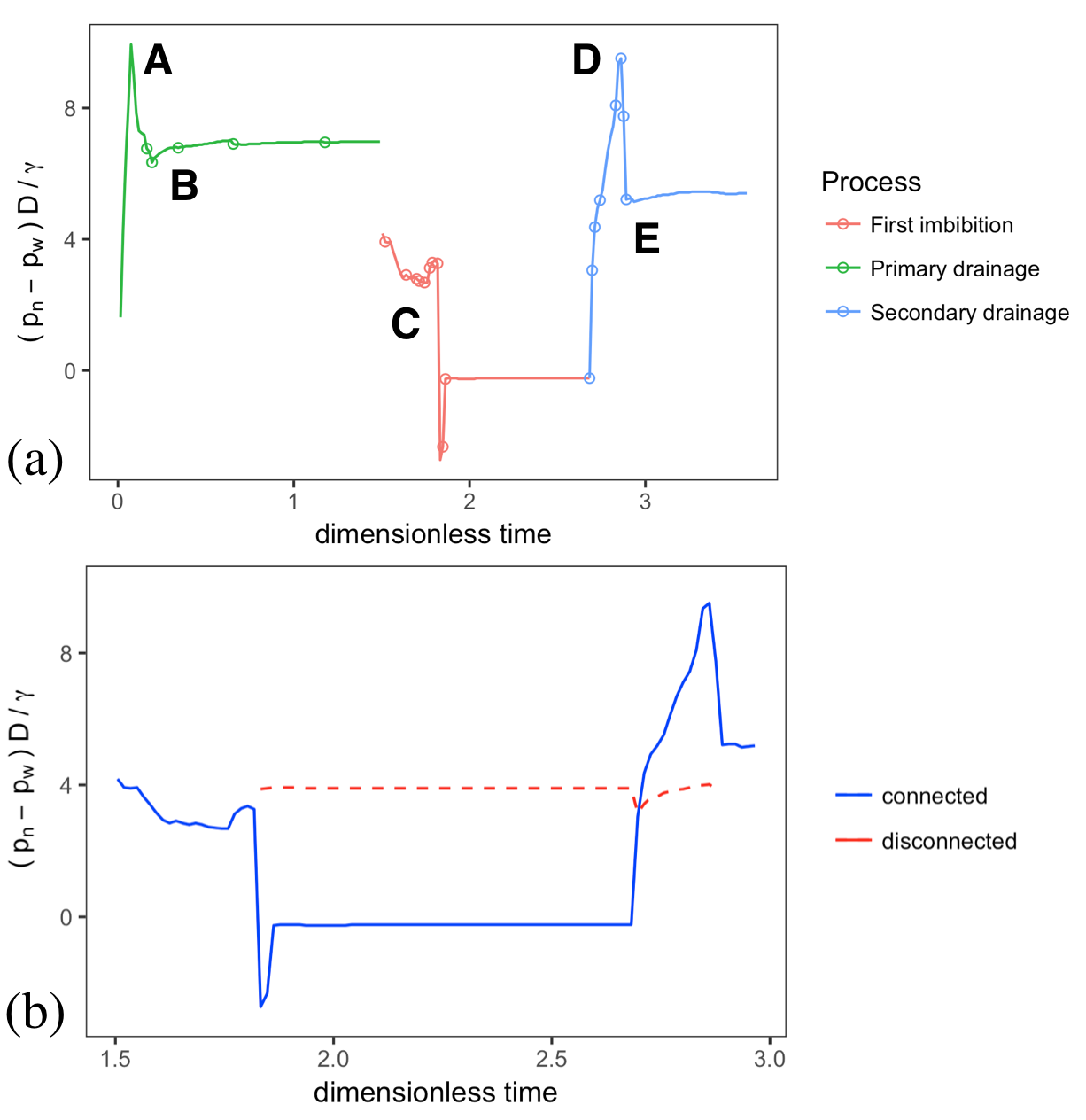}
\caption{Response of the capillary pressure to connectivity events:
(a) the difference between the pressure of the connected water and non-wetting
fluid show that non-smooth disruptions to the capillary pressure landscape align 
with connectivity events identified by changes in Euler characteristic (circles);
(b) snap-off causes large disruptions in capillary pressure, with non-wetting
fluid trapped at higher capillary pressure than the bulk fluid.}
\label{fig:pc}
\end{figure}

Of particular interest is the manner in which the discrete events captured by the Euler characteristic influence other physical variables. Fig. 
\ref{fig:Xn}
shows how the fluid pressures behave during displacement. Primary 
and secondary drainage each show a rapid spike in the pressure difference due
to the entry pressure. The entry sequence along primary drainage is 
labeled as A--C and matches the configurations from Fig. 
\ref{fig:pc}a.
The high pore entry pressure is consistent with expected behavior that the
largest energy barriers in the capillary dominated system are due to to
pore entry \cite{Berg_Ott_etal_13}. The pore entry sequence along
secondary drainage is labeled H -- J.
On secondary drainage, the entry pressure is slightly reduced
due to the presence of trapped fluid. This shows that
the first coalescence event depicted in Fig. \ref{fig:porescale}a--c corresponds
exactly to the maximum fluid pressure difference observed during pore entry. 
This is distinct from the initial pore entry observed during primary drainage, 
because it is the fluid coalescence event that causes the time derivative of
the pressures to change. In the absence of coalescence a Haines jump can be viewed 
as locally smooth process that happens on a very fast timescale. On the other hand
coalescence events are not locally smooth. Indeed, non-smooth disruptions to the
capillary pressure landscape are observed each time a coalescence event occurs. 

Like coalescence, snap-off events are associated with clear jump conditions in
the bulk fluid pressures. The snap-off sequence shown in Fig. \ref{fig:porescale} is
labeled as D--F in Fig. 
\ref{fig:pc}. As loops of fluid are destroyed, corresponding
fluctuations in the fluid pressure difference are observed. The largest disruption is associated with the snap-off event.
The capillary pressure for the connected and disconnected regions of non-wetting fluid
is shown in Fig. 
\ref{fig:pc}b. Consistent with the observation of other authors,
fluid is trapped at a higher capillary pressure than what is measured in the connected
fluid phases at the instant of snap-off \cite{Roof_1970}.
As soon as snap-off occurs, the time derivative of the  pressure difference between
the two connected fluid phases immediately jumps from strongly negative to strongly
positive. This occurs as each fluid region relaxes toward its preferred equilibrium 
state, which is possible because the two fluid regions are no 
longer mixing after the snap-off event. The final pressure for the trapped region is determined based on geometry, with the fluid assuming a minimum energy configuration within the pores where it is trapped. 

For both drainage and imbibition processes, changes in connectivity frequently 
correspond to a rapid jump in the time derivative of the pressure. 
The magnitude of the jump varies significantly, with 
events that most significantly alter the fluids ability to mix causing the largest disruptions. This is consistent with a local breakdown of the ergodic hypothesis.
Physically, the equilibrium state that the system relaxes to is immediately altered
when a connectivity change occurs. This is most evident in the snap-off process, 
because the trapped fluid relaxes toward a separate equilibrium capillary pressure,
which is only possible due to the snap-off event. While the relaxation process itself
is not instantaneous, the associated change in the time derivative is an instantaneous 
jump condition.

\begin{figure}[ht]
\centering
\includegraphics[width=1.0\linewidth]{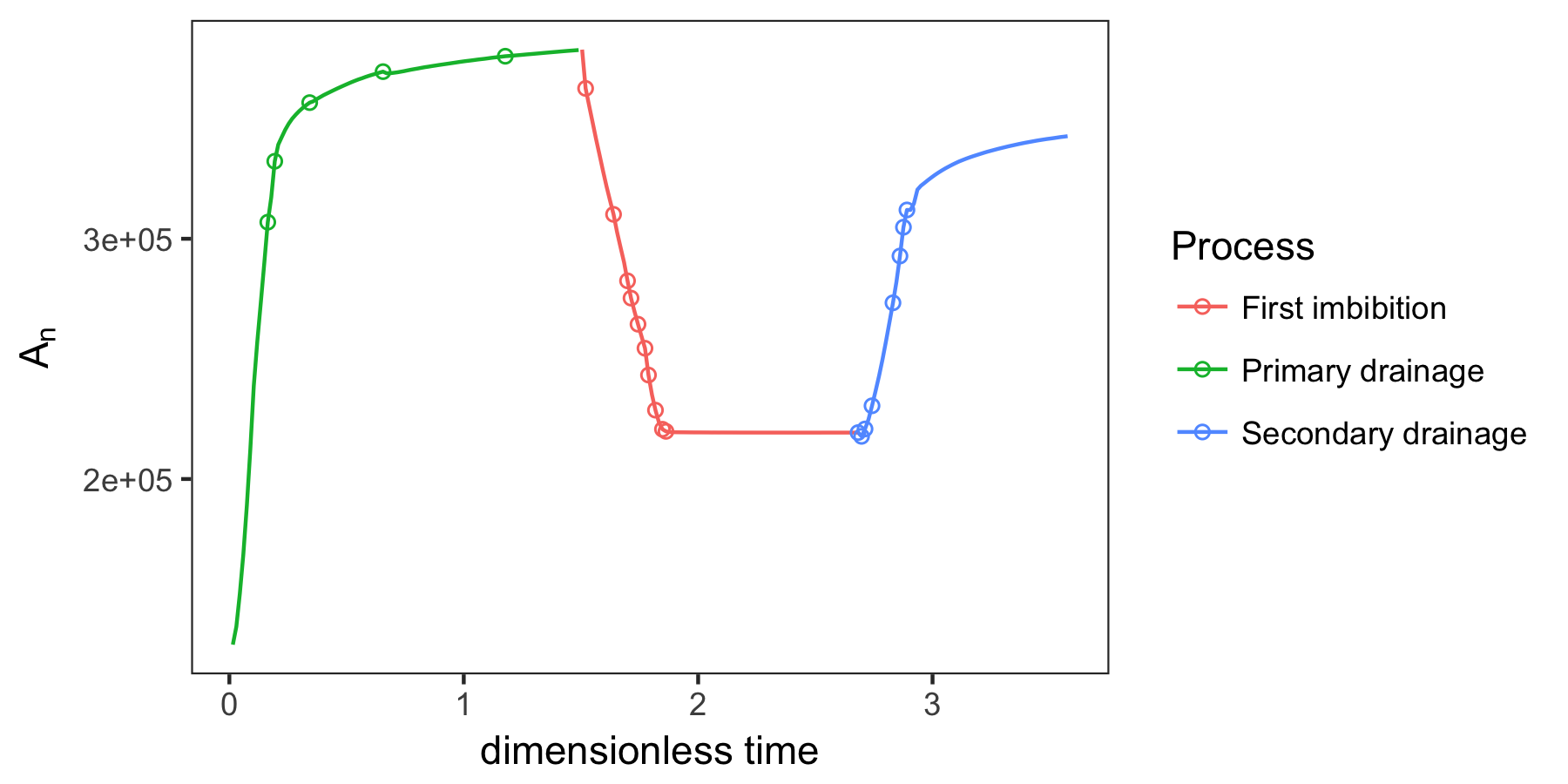}
\includegraphics[width=1.0\linewidth]{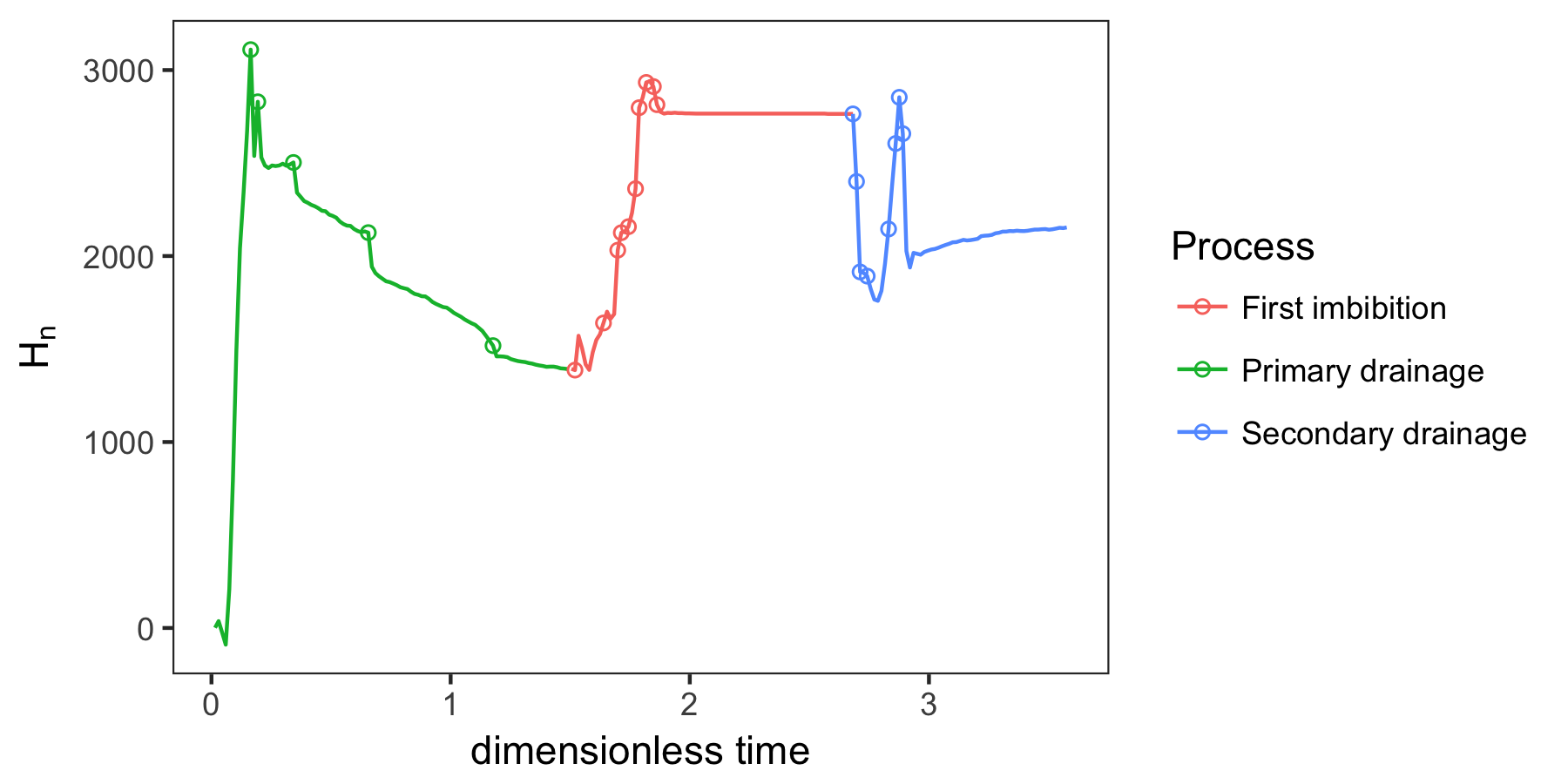}
\caption{Time evolution for the surface area and mean width. Connectivity events
cause jump conditions in the mean width, which drive discontinuous behavior in
the bulk fluid pressures.}
\label{fig:measures}
\end{figure}

The behavior of the surface area and mean width yields further insight into the effects
of the connectivity dynamics. The time evolution for these variables is shown in Fig. \ref{fig:measures}.
Since coalescence and snap-off events impact only a small fraction of
the surface area, local discontinuities have a negligible impact on the total surface area.
On the other hand, the effect of the local discontinuity on the mean curvature is significant.
Since coalescence and snap-off events occur at the fluid meniscus, a strong impact on the capillary pressure is unavoidable -- the total curvature and mean curvature are coupled. 
The implication is that the discontinuity in the capillary force
due to connectivity events is directly responsible for the corresponding behavior 
in the bulk fluid pressures. Based on Fig. \ref{fig:measures} one can clearly identify both
the discrete and continuous portions of the geometric evolution process.



\section*{Conclusions}

Inherently discontinuous aspects of geometric evolution processes arise 
due to the use of the Gibbs dividing surface in classical thermodynamic descriptions. 
Key questions arise regarding the application of continuous theory to systems that 
undergo topological changes. We classify the geometric discontinuities into eight 
categories, noting that any geometric transformations that do not
involve a topological change are continuous. The continuous and discontinuous sub-regions for geometric evolution can be chained together to describe how complex structures evolve. It is demonstrated that discontinuities resulting from topological change imply that the system
dynamics will be piece-wise continuous. Expressions are provided to describe geometric evolution,
with the topological state of the system being inferred from a state relationship 
relating the four geometric invariants. Noting that Noether's theorem is derived based on explicit assumptions of differentiability, geometric discontinuities result in symmetry-breaking when the singular points are considered. Results of classical thermodynamics link the geometric invariants with standard thermodynamic quantities. Discontinuous geometric effects can thereby propagate to 
the associated physical variables.
The consequences are demonstrated for flow of immiscible fluids in porous media,
which confirm that the time derivative for the fluid pressures is discontinuous
in the vicinity of a topological change. We show that these discontinuities are
mechanical in nature, resulting from a discontinuity in the mean curvature 
at the fluid boundary. Further work is needed to explore the consequences of discontinuous phenomena for statistical thermodynamics and continuum-mechanical
descriptions of the system behavior within this broad class of problems.
Efforts to understand how topological changes impact energy dissipation are of particular
importance. 

\subsection*{Acknowledgements}
\begin{acknowledgments}
`An award of computer time was provided by the Department of Energy Summit Early Science program. This research also used resources of the Oak Ridge Leadership Computing Facility, which is a DOE Office of Science User Facility supported under Contract DE-AC05-00OR22725. 
\end{acknowledgments}

\end{document}